# Planetary Entry Probe Dataset: Analysis and Rules of Thumb for Future Missions

Athul Pradeepkumar Girija [1],[**],[**]

[1]School of Aeronautics and Astronautics, Purdue University, West Lafayette, IN 47907, USA

## ABSTRACT

Since the beginning of robotic interplanetary exploration nearly six decades ago, successful atmospheric entry has been accomplished at Venus, Earth, Mars, Jupiter, and Titan. More entry probe missions are planned to Venus, Titan, and Uranus in the next decade. Atmospheric entry subjects the vehicle to rapid deceleration and aerothermal loads which the vehicle must be designed for, to deliver the robotic instruments inside the atmosphere. The design of planetary probes and their mission architecture is complex, and involves various engineering constraints such as peak deceleration, heating rate, heating load, and communications which must be satisfied within the budget and schedule of cost constrained mission opportunities. Engineering design data from previous entry probe missions serve as a valuable reference for designing future missions. The present study compiles an augmented version of the blue book entry probe dataset, performs a comparative analysis of the entry conditions, and provides engineering rules of thumb for design of future missions. Using the dataset, the present study proposes a new empirical correlation which aims to more accurately predict the thermal protection system mass fraction for high heat load conditions during entry and aerocapture at Uranus and Neptune.

[****] To whom correspondence should be addressed, E-mail: athulpg007@gmail.com

## I. INTRODUCTION

Entry probes are a unique class of space vehicles which are used to deliver robotic instruments inside planetary atmospheres [1]. These include atmospheric descent vehicles such as Pioneer Venus and Galileo probes, landers such as Venera and Huygens, rovers such as Mars 2020 and Zhurong, and balloons such as Vega-1. These robotic missions have returned invaluable in-situ measurements that simply cannot be obtained by remote sensing, and have been a critical element in expanding our knowledge and understanding of the Solar System [2]. For planetary bodies with atmospheres, an atmospheric entry system is required to deliver the scientific instruments inside the atmosphere or to the surface where it can perform in situ measurements. Since the beginning of robotic space exploration, successful atmospheric entry has been accomplished at Venus, Earth, Mars, Jupiter, and Titan in the past six decades. More entry probe missions are planned to Venus (DAVINCI), Titan (Dragonfly), and Uranus (Flagship Mission) in the next decade (2030-2040). Atmospheric entry subjects the vehicle to rapid deceleration and aero-thermal loads which the vehicle must be designed for, to deliver a robotic instrument inside the atmosphere. The entry conditions strongly depend on the destination's gravity, atmospheric structure, and chemical composition [3]. The design of planetary probes and their mission architecture is complex, and involves various engineering constraints such as peak deceleration, heating rate, heating load, and communications which must be satisfied within the budget and schedule of cost-constrained mission opportunities. Engineering design data from previous entry probe missions serve as a valuable reference for designing future missions. For example, a system engineer may be interested in quickly finding out the atmospheric entry conditions (entry-flight path angle, speed etc.) for a probe at Venus or Titan, and roughly estimating the expected conditions for a probe under these conditions. NASA's Ames Research Center has compiled a 'Probe Blue Book' as a ready reference manual for system engineers to address this need [4]. The most recent edition of book provides a list of 32 unmanned entry probe missions with their entry conditions, vehicle design, and estimated or actual entry conditions. Of particular interest during the conceptual design phase are key design variables such the peak deceleration, peak heat rate, total heat load, and the thermal protection system (TPS) mass fraction to analyze the preliminary feasibility of the mission architecture [5]. Planetary entry carpet plots are another method for quickly analyzing the design trade space [6]. There also exists tools such as the Aerocapture Mission Analysis Tool (AMAT) which can quickly simulate a given planetary entry and estimate the expected conditions such as deceleration and heating loads [7, 8]. The present study compiles an augmented version of the blue book entry probe dataset, performs a comparative analysis of the entry conditions, and provides engineering rules of thumb for designing future missions.

## II. DATASET DESCRIPTION

The dataset is built upon the 32 missions listed probe blue book, which is augmented with data for the Venera entry probes from Dutta et al. [9], recent missions such as Tianwen-1 [10], planned missions such as Dragonfly [11], and some proposed entry and aerocapture mission concepts which have been not flown yet [12]. The total number of missions in the dataset is n = 53. The data fields are: Source, Mission, Planet, Launch, Year (Entry), entry flight path angle (EFPA), planet-relative entry speed, trim L/D, control method, ballistic coefficient (beta), peak deceleration, vehicle shape, nose radius, base area, base diameter, entry mass, science instrument payload mass, TPS mass fraction, peak convective heating rate, peak radiative heating rate, peak heating stagnation pressure, integrated heat load, peak total stagnation point heat rate, forebody TPS material, forebody TPS thickness, forebody TPS mass, aftbody TPS material, aftbody TPS thickness, aftbody TPS mass, Country, Successful (Yes or No), and Status (Flown, Planned, Concept). The dataset includes some early unmanned Apollo test flights at Earth, but does not include manned entry missions. The full dataset is available as csv file, refer to the Data Availability section to access the dataset.

## III. DATA ANALYSIS

Figure 1 shows the cumulative entry missions count from the dataset as a function of the year. The data indicates number entry missions continues to grow, but at a slower rate than that compared to the 1970s and 1980s. The spike in the late 1970s and early 1980s is mainly attributed to the numerous Venera probes and the Pioneer Venus multi-probes [13]. A quadratic curve fit (where x = years since 1965) appears to reasonably model the number of entry missions.

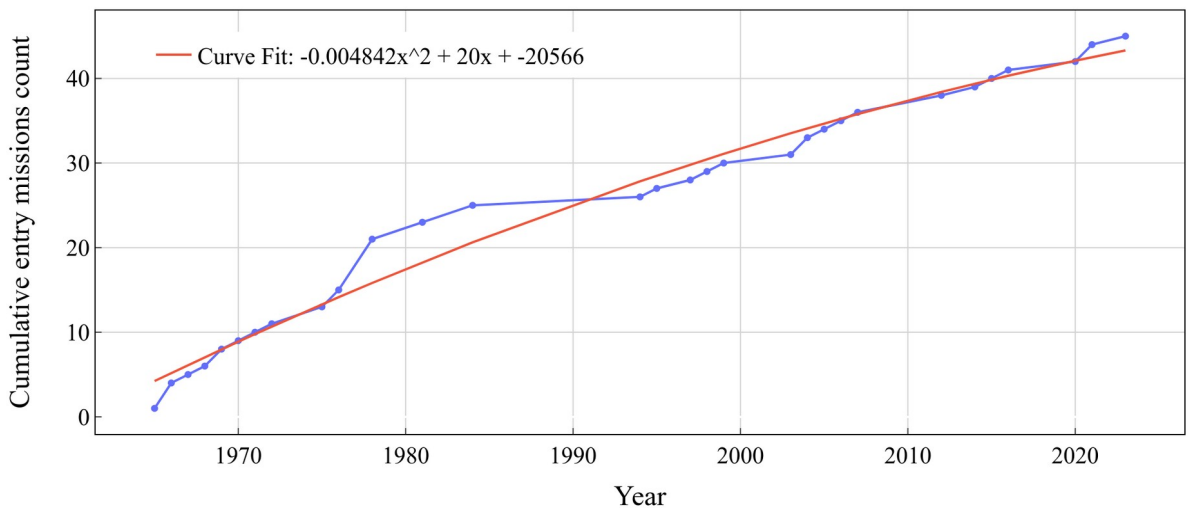

Figure 1. Cumulative entry missions count from the dataset.

Figure 2 shows the number of entry missions per decade, colored by the destination which shows the great interest in Venus till the mid-1980s. After 1990, there were no Venus entry missions, and the number of probe missions have remained at about 5 per decade, half of what it was in the 1970s. In 1995, the Galileo probe accomplished the first entry at Jupiter [14]. In 2005, Huygens accomplished the first entry at Titan [15]. Figure 3 shows the count the number of entry missions. Earth, Venus, and Mars have the most missions. Despite Mars being the popular target in the last few decades, Venus still has more entry probes to date than Mars, attributed to the numerous Venera probes.

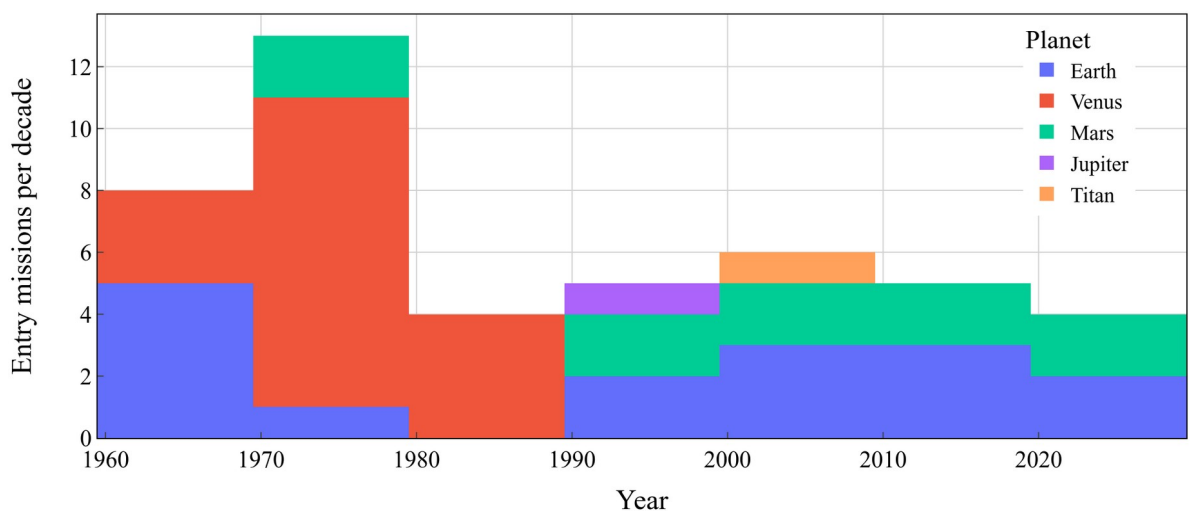

Figure 2. Number of entry missions per decade, colored by destination.

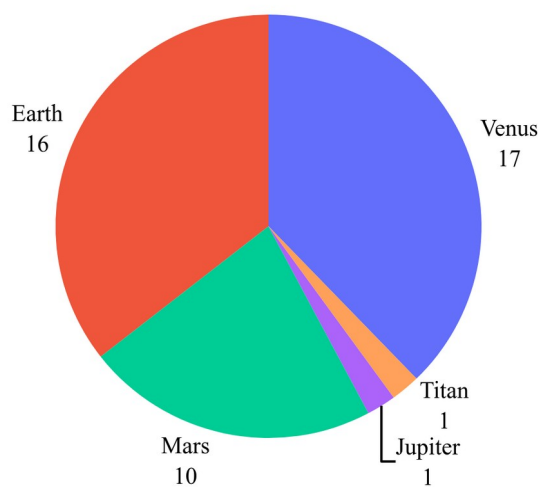

Figure 3. Number of entry missions per target in the dataset.

Figure 4 shows the number of entry missions per country in the dataset. The United States with NASA is the leading country in terms of planetary entry missions, followed by the former USSR and ESA. ESA has attempted multiple landings at Mars, but has not been successful yet. The Japanese Space Exploration Agency (JAXA) has an exemplary record with the Hayabusa series of sample return missions [16]. Figure 5 shows the number of entry missions per decade, colored by country which highlights the Soviet Venera probes until the 1980s. The Chinese Space Agency (CNSA), achieved their first successful Mars landing with the Tianwen-1 lander in 2021 [17].

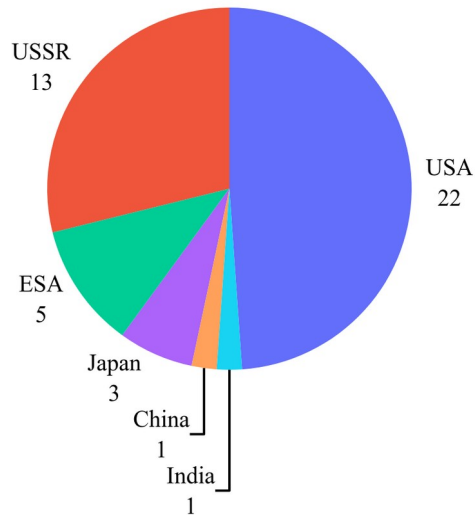

Figure 4. Number of entry missions per country in the dataset.

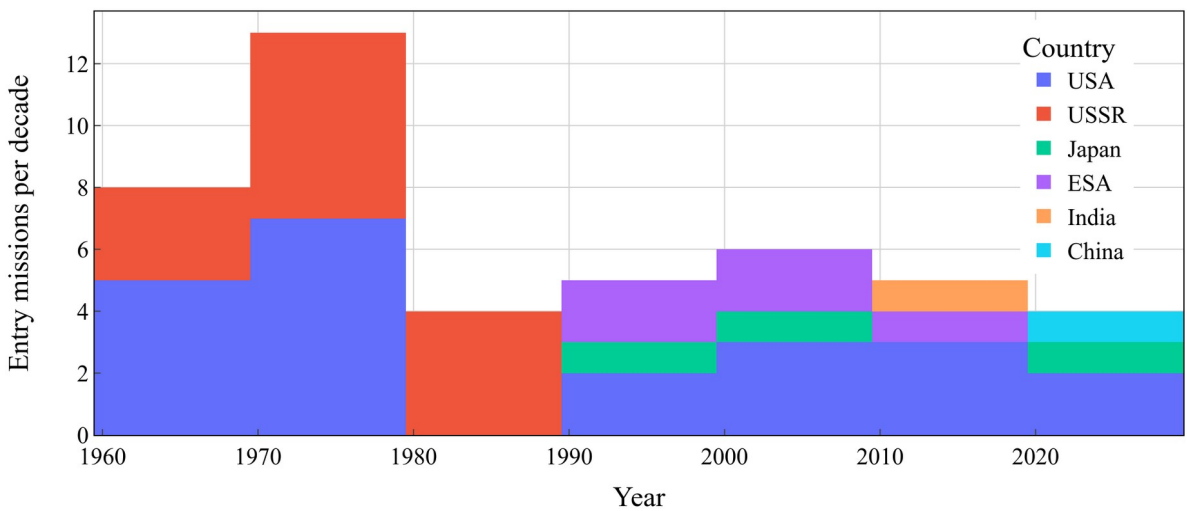

Figure 5. Number of entry missions per decade, colored by country.

Figure 6 shows the entry conditions (speed and EFPA) for Venus missions from the dataset. The early Venus missions chose a steep EFPA to minimize the total heat load and TPS mass, but sustained very high deceleration and heating loads. Venera 4 entered the steepest, and sustained the highest deceleration load ever recorded for a probe (500g). The four Pioneer Venus probes entered at various flight-path angles due to their mission requirements. Figure 7 shows the entry conditions (speed and EFPA) for unmanned Apollo entry and Earth sample return missions from the dataset. Compared to the steep entries at Venus, Earth entries have all been relatively shallow at [-14, -8] degrees. The low speed entries are suborbital test flights, while the high speed entries are from Earth sample return missions.

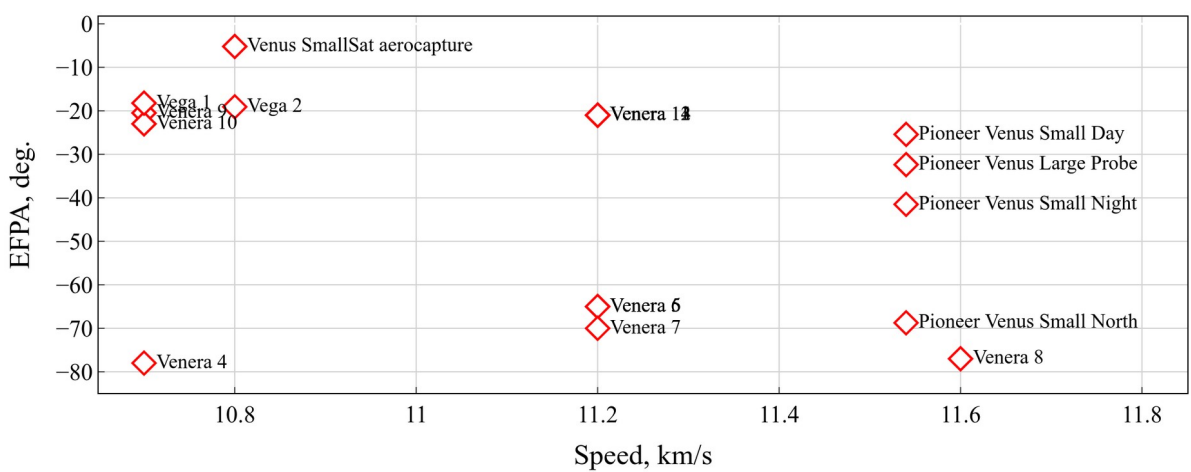

Figure 6. Venus entry conditions (speed and EFPA) from the dataset.

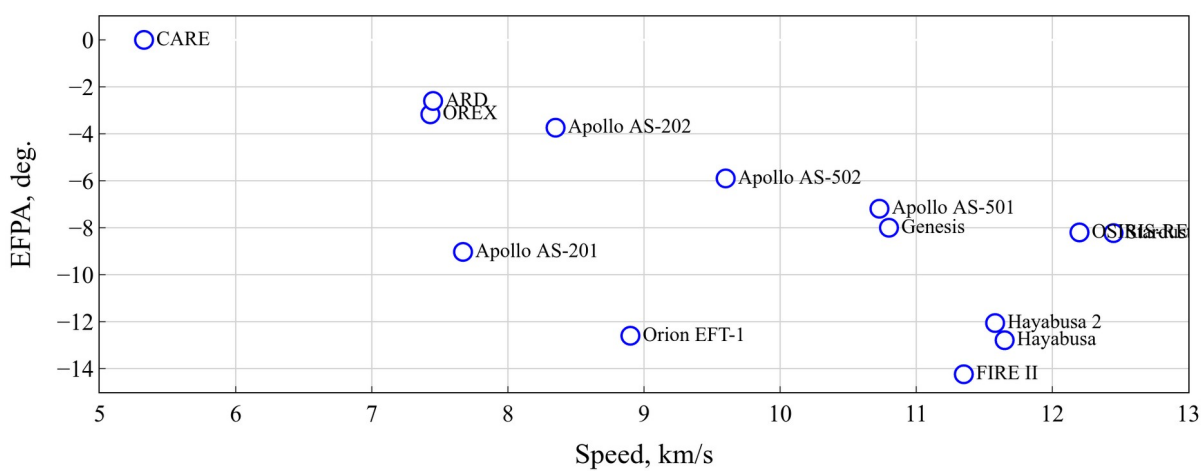

Figure 7. Earth entry conditions (speed and EFPA) from the dataset.

Figure 8 shows the entry conditions (speed and EFPA) for Mars missions from the dataset. Compared to Venus and Earth, Mars with its lower gravity enables lower entry speeds in the range of 4.5–7.5 km/s, and EFPA in the range of [-18, -11] degrees. The thin Martian atmosphere presents relatively benign aero-thermal conditions compared to Venus or Earth, but also poses challenges as the atmosphere simply does not slow down the vehicle enough.

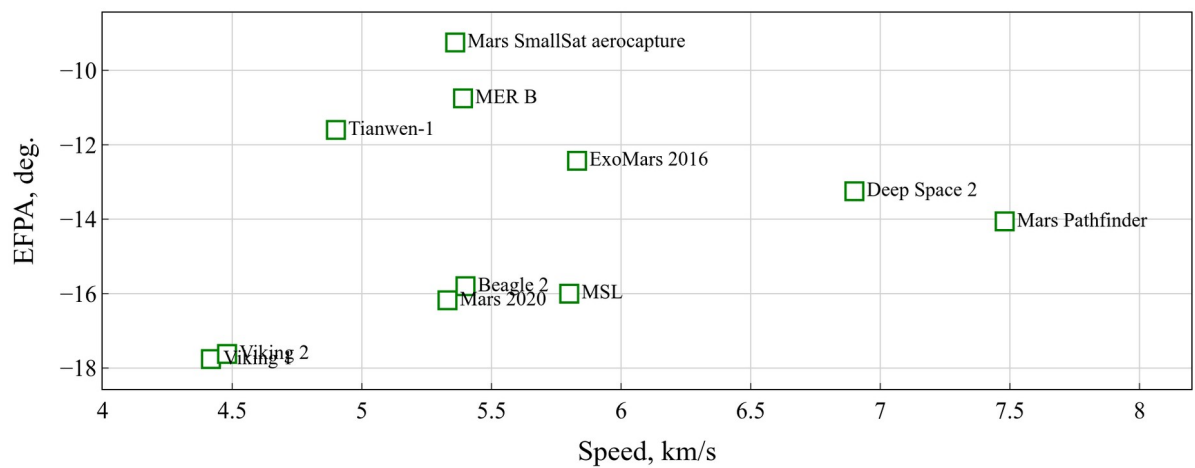

Figure 8. Mars entry conditions (speed and EFPA) from the dataset.

Figure 9 shows the combined plot of all entry conditions from the dataset, colored by planet. Jupiter with its enormous gravity results in extremely large entry speeds, with the Galileo probe entering at 47.4 km/s. Uranus and Neptune also result in large entry speeds (20–30 km/s). Titan with its thick extended atmosphere and low gravity requires a steep entry in the range of [-65, -50] deg., but the aero-thermal conditions are still comparable to Mars.

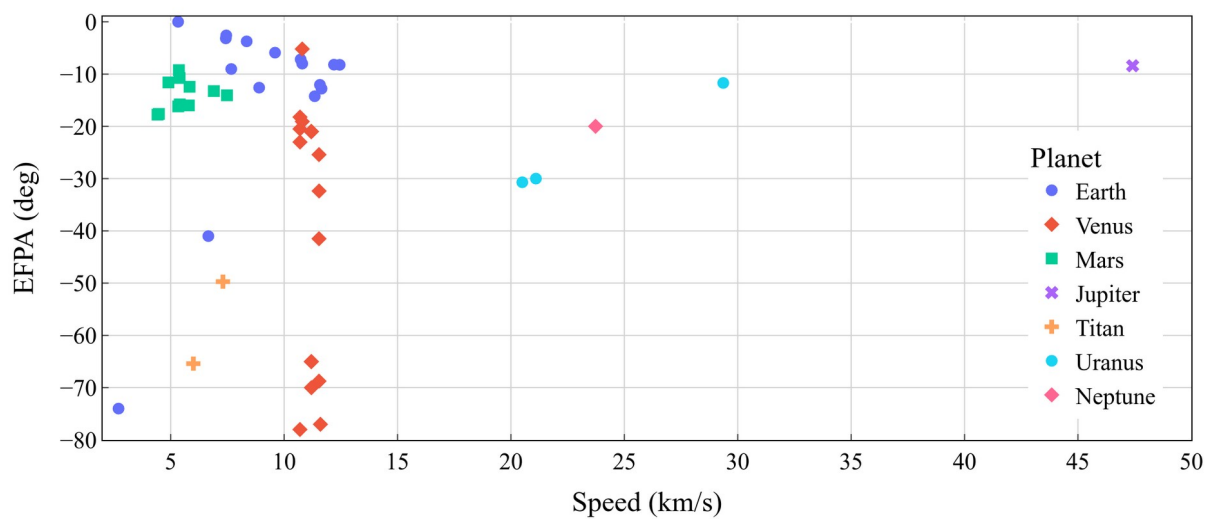

Figure 9. Combined plot showing all the entry conditions from the dataset.

Figure 10 shows the distribution of ballistic coefficient and entry mass in the dataset. The high mass points at Earth are Apollo and Orion flight test articles which weigh in the thousands of kg. The two heavy Mars entries (3200 kg) are the MSL and the Mars 2020 entry vehicles. Other entry probes are much lighter. A typical outer planet atmospheric probe such as the Galileo weighs about 250–320 kg and carries a descent probe which accounts for roughly half the entry mass. Figure 11 shows the distribution of peak deceleration and peak stagnation point heat rate. The Venera 4 probe which entered the steepest at -80 deg. had the highest combination of deceleration and heating.

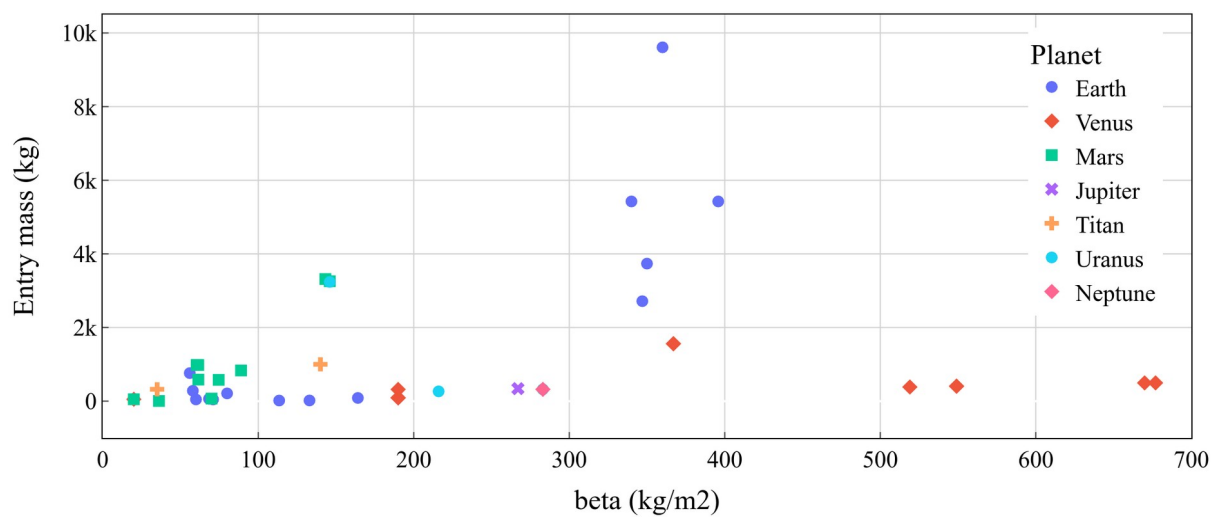

Figure 10. Distribution of ballistic coefficient and entry mass.

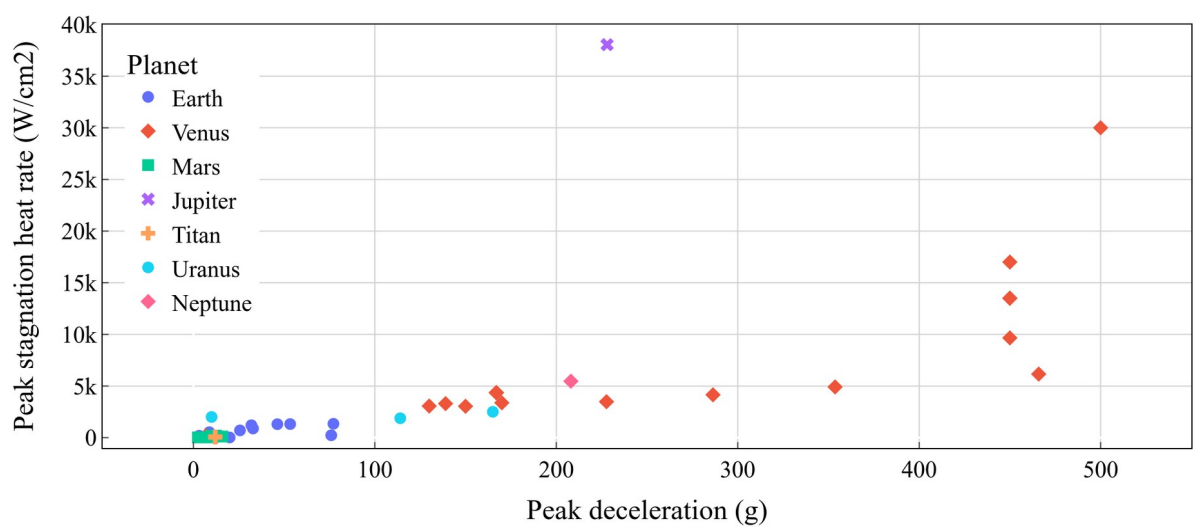

Figure 11. Peak deceleration and stagnation point heat rate distribution.

Figure 12 shows the peak deceleration and stagnation point heat rate for Earth, Mars, and Titan. Compared to Earth, Mars and Titan are both relatively benign targets for deceleration and aero-heating. Mars and Titan entries have decelerations under 12g, and stagnation point heating rates under 200 W/cm². Earth sample return missions on the other hand can encounter much higher decelerations (30-50g) and heat rates up to 1400 W/cm². Figure 13 shows the distribution of heat rate and stagnation pressure. Both Venus and the outer planets, Uranus and Neptune present a particularly challenging environment in terms of the combination of heat rate and stagnation pressure.

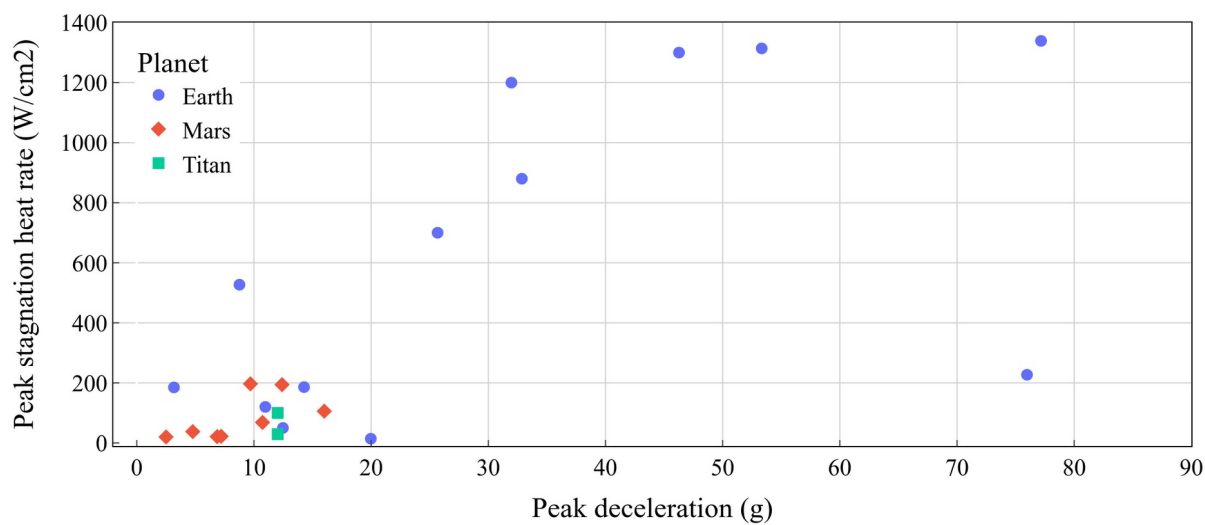

Figure 12. Peak deceleration and stagnation point heat rate for Earth, Mars, Titan.

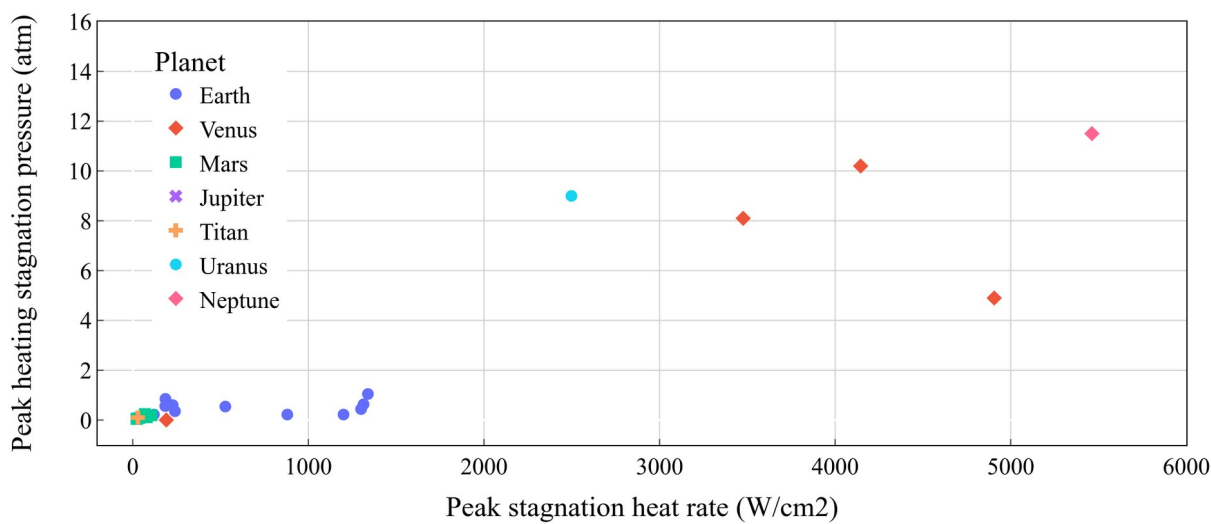

Figure 13. Peak stagnation point heat rate and stagnation pressure.

Figure 14 shows the peak heat rate and stagnation pressure for Earth, Mars, and Titan. Compared to Venus and the outer planets, these destinations are quite benign in terms of the combination of heat rate and stagnation pressure.

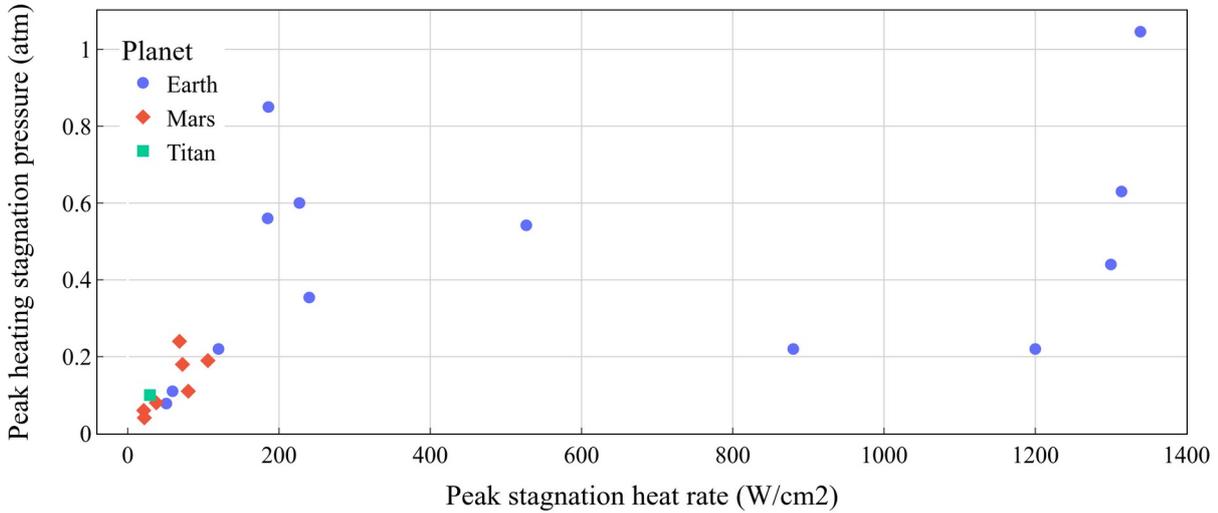

Figure 14. Peak heat rate and stagnation pressure for Earth, Mars, and Titan.

Figure 15 shows the distribution of the integrated total heat load and the TPS mass fraction. Previously in the literature, a legacy empirical relationship TPS% = $0.091Q^{0.51575}$ (Q = heat load in J/cm2) was derived by Laub and Venkatapathy following a similar data-driven analysis [18]. However, the data used for correlation relied on the legacy Carbon-phenolic and over-predicts the TPS mass fraction when applied to the entry probes which uses HEEET.

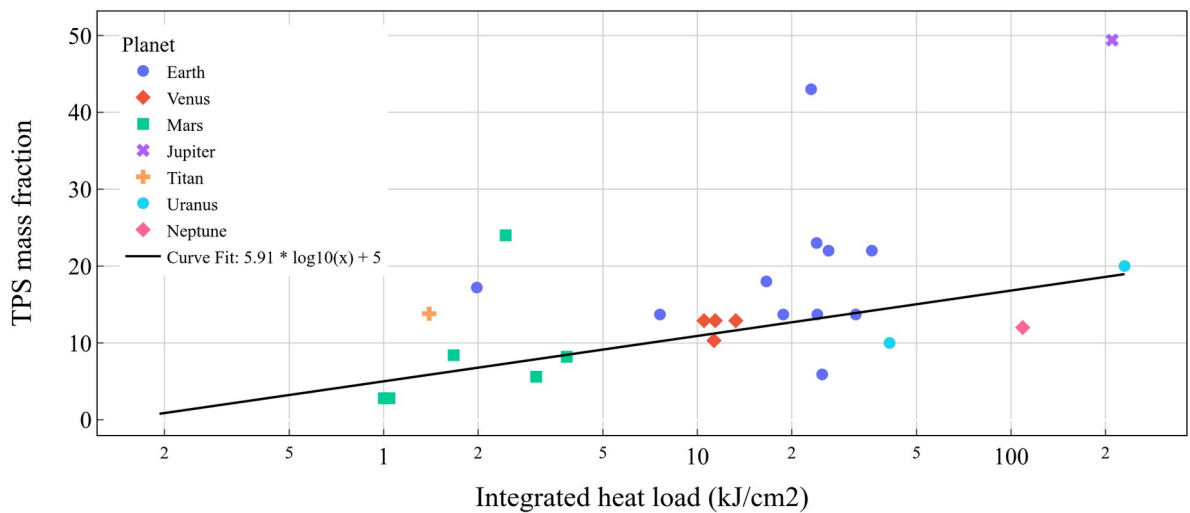

Figure 15. Total heat load and TPS mass fraction.

Using the dataset, the present study proposes a new empirical correlation which aims to more accurately predict the TPS mass fraction for entry probes using HEEET, especially at high heat loads such as those at ice giants [19]. The new correlation is TPS% = 5.91 * $\log_{10}Q$ + 5, where Q = heat load in kJ/cm2. Note the heat load is assumed in kJ/cm2. Table 1 compares the two correlations for Uranus and Neptune probes from the Ice Giant Pre-Decadal Study and an estimate from a Uranus aerocapture study. The legacy correlation over-predicts the TPS mass fraction significantly, while the proposed new correlation is more accurate even though it still overestimates by about 5% in some cases.

Table 1. Comparison of legacy and proposed TPS% correlation.

| Mission | Planet | Source | Heat load, kJ/cm2 | TPS% | Legacy correlation | New correlation |
|---|---|---|---|---|---|---|
| Probe | Uranus | Elliot et al. (2017) | 41 | 10 | 21 | 14.5 |
| Probe | Neptune | Elliot et al. (2017) | 109 | 12 | 36 | 17.0 |
| Aerocapture | Uranus | Girija et al. (2023) | 230 | 20 | 53 | 18.9 |

## IV. RULES OF THUMB FOR FUTURE ENTRY AND AEROCAPTURE MISSIONS

The high deceleration and aero-thermal loads at Venus present significantly more challenging entry conditions than Earth or Mars. This implies probe entry at Venus continues to be a difficult challenge. This also makes aerocapture using heritage blunt-body aeroshells not compelling as the heating rates are quite high and the TPS mass fraction is substantial [20]. However, aerocapture is still a viable option for inserting small satellites into Venus orbit using low-ballistic coefficient entry systems [21]. By decelerating much higher up in the thinner atmosphere the deployable systems can keep the heating rates much lower than rigid aeroshells. Compared to rigid aeroshells which encounter 100s of g and 1000s of W/cm$^2$, low-ballistic coefficient systems can achieve 50-60g, and a few 100 W/cm$^2$. These systems have potential applications for delivering probes and landers [22], delivering standalone small satellites to orbit using aerocapture [23], small satellites as part of larger mission [24], and atmospheric sample return missions [25]. Sample return missions at Earth using rigid aeroshells encounter heating rates as high as 1400 W/cm$^2$ [26].

With the lowest deceleration and aero-thermal loads in the inner Solar System, Mars presents a relatively benign target for both entry probes delivering landers and aerocapture [27, 28]. The deceleration loads are under 12g, and the heating rates are under 200 W/cm$^2$. This makes Mars an ideal destination for deployable entry systems such as ADEPT for delivering small landers. and orbiters, and a low-cost aerocapture technology demonstrator mission [29].

Jupiter's enormous gravity well results in very high entry speeds, and results in the most extreme deceleration and aero-thermal loads in the Solar System. With entry speeds as high as 50 km/s, the probe encounters 100s of g, and several 10,000 W/cm$^2$ of heating rate, Jupiter presents severe challenges for atmospheric entry and aerocapture [30].

Titan's low gravity and thick extended atmosphere makes it a unique destination for atmospheric entry. The low gravity results in low entry speeds, and the thick extended atmosphere slowly decelerates the probe, encountering decelerations of about 12g and a peak heating under 200 W/cm$^2$ which makes it comparable to Mars. The relatively benign entry conditions make it an ideal candidate for a future Titan orbiter mission using aerocapture [31].

At the ice giants, heating rates are in the range of 1000-5000 W/cm$^2$, which falls within the limits of HEEET TPS. While not considered for the baseline Uranus mission concepts [32, 33], aerocapture can considerably reduce the flight time, as well as increase payload mass for future ice giant missions [34, 35]. The proposed new correlation can be used to more accurately estimate the HEEET TPS mass fraction for ice giant probe entry and aerocapture [36].

## V. CONCLUSIONS

The design of planetary probes and their mission architecture is a complex endeavor, and involves various engineering constraints such as peak deceleration, heating rate, heating load, and communications which must be satisfied within the budget and schedule of cost-constrained mission opportunities. Engineering design data from previous entry probe missions serve as a valuable reference for designing future missions. The present study compiled an augmented version of the blue book entry probe dataset, performed a comparative analysis of the entry conditions (speed and EFPA), and the expected deceleration and aero-thermal conditions at various destinations across the Solar System. The high deceleration and aero-thermal loads at Venus present significantly more challenging entry conditions than the Earth or Mars, but low ballistic coefficient systems can alleviate this problem to some extent. Sample return missions at Earth using rigid aeroshells encounter heating rates as high as 1400 W/cm$^2$. Mars presents a relatively benign target for both entry probes delivering landers and aerocapture. With entry speeds as high as 50 km/s, the probe encounters 100s of g, and several 10,000 W/cm$^2$ of heating rate, Jupiter presents severe challenges for atmospheric probe entry. At Titan, the probe experiences decelerations of about 12g and a peak heating under 200 W/cm$^2$ which makes it comparable to Mars. At the ice giants, heating rates are in the range of 1000-5000 W/cm$^2$, and heat loads are in the range of 100–200 kJ/cm2. The high heat loads at the ice giants present a challenge for aerocapture, but the TPS mass fraction can still be reasonable with HEEET. Using the dataset, the study proposed a new empirical correlation which estimates the TPS mass fraction with better accuracy than the legacy correlation, especially for high heat loads.

## DATA AVAILABILITY

The dataset is available at http://dx.doi.org/10.13140/RG.2.2.17150.56641. The dataset is not an exhaustive list, has some gaps in the data, and some data points which were estimated by the author. The proposed TPS% correlation must be used as a preliminary engineering estimate, and needs to be verified by higher fidelity analysis. The source code used to make the charts in the paper will be made available by the author upon request.

## REFERENCES

[1] Ball A. et al.,"Planetary Landers and Entry Probes," Cambridge University Press, 2007.

[2] Beatty, J. Kelly et al., "The New Solar System," Cambridge University Press, 1999.

[3] Girija AP, "Comparative Study of Planetary Atmospheres and Implications for Atmospheric Entry Missions," arXiv, Vol. 2307, No. 16277, 2023, pp. 1-15
https://doi.org/10.48550/arXiv.2307.16277

[4] NASA Ames Research Center, Entry Systems and Technology Division. Planetary Mission Entry Vehicle Quick Reference Guide. NASA, 2022. NASA/SP-20220010761.

[5] Girija AP et al., "A Unified Framework for Aerocapture Systems Analysis," *AAS/AIAA Astrodynamics Specialist Conference,* 2019, pp 1-21.
https://doi.org/10.31224/osf.io/xtacw

[6] Girija AP, "A Survey of the Design Trade Space for Atmospheric Entry, Descent, and Landing Missions," arXiv, Vol. 2308, No. 03238, 2023, pp. 1-11
https://doi.org/10.48550/arXiv.2308.03238

[7] Girija AP, "A Systems Framework and Analysis Tool for Rapid Conceptual Design of Aerocapture Missions," Ph.D. Dissertation, Purdue University Graduate School, 2021.
https://doi.org/10.25394/PGS.14903349.v1

[8] Girija AP et al. "AMAT: A Python package for rapid conceptual design of aerocapture and atmospheric Entry, Descent, and Landing (EDL) missions in a Jupyter environment," *Journal of Open Source Software,* Vol. 6, No. 67, 2021, pp. 3710.
https://doi.org/10.21105/joss.03710

[9] Dutta S. et al., "Mission Sizing and Trade Studies for Low Ballistic Coefficient Entry Systems to Venus," *IEEE Aerospace Conference,* 2012, pp. 1-14. https://doi.org/10.1109/AERO.2012.6187002

[10] Huang X. et al., "The Tianwen-1 guidance, navigation, and control for Mars entry, descent, and landing," *Space: Science & Technology,* Vol. 2021, 2021. https://doi.org/10.34133/2021/9846185

[11] Barnes, JW et al. "Science goals and objectives for the Dragonfly Titan rotorcraft relocatable lander," *The Planetary Science Journal,* Vol. 2, No. 4, 2021, pp. 130. https://doi.org/10.3847/PSJ/abfdcf

[12] Girija AP, "A Flagship-class Uranus Orbiter and Probe mission concept using aerocapture," *Acta Astronautica* Vol. 202, 2023, pp. 104-118.
https://doi.org/10.1016/j.actaastro.2022.10.005

[13] Marov M. et al., "The Planet Venus," Yale University Press, 1998.

[14] Young RE, "The Galileo probe: how it has changed our understanding of Jupiter," *New Astronomy Reviews,* Vol. 47, No. 1, 2003, pp. 1-51. https://doi.org/10.1016/S1387-6473(02)00272-5

[15] Lebreton JP, et al., "An overview of the descent and landing of the Huygens probe on Titan," *Nature,* Vol. 438, No. 7069, 2005, pp. 758-764. https://doi.org/10.1038/nature04347

[16] Kawaguchi J. et al., "Hayabusa—Its technology and science accomplishment summary and Hayabusa-2," *Acta Astronautica,* Vol. 62. No. 10-11, 2008, pp. 639-647.
https://doi.org/10.1016/j.actaastro.2008.01.028

[17] Sun Z, and Rao W, "Entry, Descent, and Landing of China's Tianwen-1 Mars Mission," *Space: Science & Technology,* Vol. 2022, 2022. https://doi.org/10.34133/2022/9809054

[18] Laub B and Venkatapathy E, "Thermal protection system technology and facility needs for demanding future planetary missions," *Proceedings of the International Workshop Planetary Probe Atmospheric Entry and Descent Trajectory Analysis and Science,* 2004.
https://adsabs.harvard.edu/full/2004ESASP.544..239L

[19] Girija AP, et al., "Feasibility and performance analysis of neptune aerocapture using heritage blunt-body aeroshells," *Journal of Spacecraft and Rockets,* Vol. 57, No. 6, 2020, pp. 1186-1203.
https://doi.org/10.2514/1.A34719

[20] Girija AP, Lu Y, and Saikia SJ, "Feasibility and mass-benefit analysis of aerocapture for missions to Venus," *Journal of Spacecraft and Rockets,* Vol. 57, No. 1, 2020, pp. 58-73.
https://doi.org/10.2514/1.A34529

[21] Girija AP, Saikia SJ, and Longuski JM, "Aerocapture: Enabling Small Spacecraft Direct Access to Low-Circular Orbits for Planetary Constellations," *Aerospace,* Vol. 10, No. 3, 2023, pp. 271.
https://doi.org/10.3390/aerospace10030271

[22] Wercinski, P, "Adaptable Deployable Entry and Placement Technology (ADEPT) Enabling Future Science Missions," *NASA Tech Showcase,* 2023.
https://ntrs.nasa.gov/citations/20220019227

[23] Austin A et al., "Enabling and Enhancing Science Exploration Across the Solar System: Aerocapture Technology for SmallSat to Flagship Missions," *Bulletin of the American Astronomical Society,* Vol. 53, No. 4, 2021, pp. 057.
https://doi.org/10.3847/25c2cfeb.4b23741d

[24] Limaye SS et al., "Venus observing system," *Bulletin of the American Astronomical Society,* Vol. 53, No. 4, 2021, pp. 370.
https://doi.org/10.3847/25c2cfeb.7e1b0bf9

[25] Shibata E et al., "A Venus Atmosphere Sample Return Mission Concept: Feasibility and Technology Requirements," *Planetary Science Vision 2050 Workshop,* 2017, pp. 8164.

[26] Desai PN, et al. "Entry, descent, and landing operations analysis for the Stardust entry capsule," *Journal of Spacecraft and Rockets,* Vol. 45, No. 6, 2008, pp. 1262-1268.
http://dx.doi.org/10.2514/1.37090

[27] Girija AP et al. "Quantitative assessment of aerocapture and applications to future solar system exploration." *Journal of Spacecraft and Rockets,* Vol. 59, No. 4, 2022, pp. 1074-1095.
https://doi.org/10.2514/1.A35214

[28] Girija AP, "Aerocapture: A Historical Review and Bibliometric Data Analysis from 1980 to 2023," arXiv*,* Vol. 2307, No. 01437, 2023, pp 1-19.
https://doi.org/10.48550/arXiv.2307.01437

[29] Girija AP, "A Low Cost Mars Aerocapture Technology Demonstrator," arXiv*,* Vol. 2307, No. 11378, 2023, pp. 1-14.
https://doi.org/10.48550/arXiv.2307.11378

[30] Spilker T et al., "Qualitative assessment of aerocapture and applications to future missions," *Journal of Spacecraft and Rockets,* Vol. 56, No. 2, 2019, pp. 536-545.
https://doi.org/10.2514/1.A34056

[31] Girija AP, "ADEPT Drag Modulation Aerocapture: Applications for Future Titan Exploration," arXiv, Vol. 2306, No. 10412, 2023, pp 1-27.
https://doi.org/10.48550/arXiv.2306.10412

[32] Jarmak S et al., "QUEST: A New Frontiers Uranus orbiter mission concept study," *Acta Astronautica,* Vol. 170, 2020, pp. 6-26.
https://doi.org/10.1016/j.actaastro.2020.01.030

[33] Cohen I et al., "New Frontiers-class Uranus Orbiter: Exploring the feasibility of achieving multidisciplinary science with a mid-scale mission," *Bulletin of the American Astronomical Society,* Vol. 53, No. 4, 2021, pp. 323.
https://doi.org/10.3847/25c2cfeb.262fe20d

[34] Iorio L et al., "One EURO for Uranus: the Elliptical Uranian Relativity Orbiter mission." *Monthly Notices of the Royal Astronomical Society*, Vol. 523, No. 3, 2023, pp. 3595-3614
https://doi.org/10.1093/mnras/stad1446

[35] Dutta S et al., "Aerocapture as an Enhancing Option for Ice Giants Missions," *Bulletin of the American Astronomical Society,* Vol. 53, No. 4, 2021, pp. 046.
https://doi.org/10.3847/25c2cfeb.e8e49d0e

[36] Venkatapathy E. et al., "Entry system technology readiness for ice-giant probe missions," *Space Science Reviews* Vol. 216, No. 22, 2020, pp. 1-21.
https://doi.org/10.1007/s11214-020-0638-2